\documentclass[aps,prb,twocolumn,floatfix,showpacs,10pt]{revtex4-1}

\usepackage{latexsym}
\usepackage{graphicx}
\usepackage{times,psfrag}
\usepackage{amsmath}
\usepackage{dcolumn}
\usepackage{color}
\usepackage{amssymb,bm,euscript}
\usepackage{amsfonts,hyperref}
\usepackage{wasysym}
\usepackage{subfigure}

\begin{document}
\title{Magnetic orders and topological phases from $f$-$d$ exchange in pyrochlore iridates}

\author{Gang Chen}
\author{Michael Hermele}
\affiliation{Physics Department, University of Colorado, Boulder, CO 80309, USA}

\begin{abstract}
We study theoretically the effects of $f$-$d$ magnetic exchange interaction in the R$_2$Ir$_2$O$_7$ pyrochlore iridates.  
The R$^{3+}$ $f$-electrons  form localized Kramers or non-Kramers doublets, while the Ir$^{4+}$ 
$d$-electrons are more itinerant and feel a strong spin-orbit coupling.  We construct and analyze a minimal model capturing 
this physics, treating the Ir subsystem using a Hubbard-type model.  First neglecting the Hubbard interaction, 
we find Weyl semi-metal and Axion insulator phases induced by the $f$-$d$ exchange. Next, we find that $f$-$d$ exchange 
can cooperate with the Hubbard interaction to stabilize the Weyl semi-metal over a larger region of parameter space than 
when it is induced by $d$-electron correlations alone.  Applications to  experiments are discussed.
\end{abstract}

\date{\today}


\maketitle


The discovery of  time-reversal invariant topological band insulators has opened new terrain in the study of 
topological states of matter\cite{Kane10,Qi10,Hasan11}.  As spin-orbit coupling (SOC) is essential to realize 
some topological states, and given the propensity of electron correlations to induce a variety of novel phenomena, 
there is now intense interest in $5d$ transition metal oxides with simultaneously strong SOC and intermediate 
electron correlation. The pyrochlore iridates R$_2$Ir$_2$O$_7$, where R is a lanthanide (or Y),  have been proposed 
to host various topological phases.  Topological band insulators,  topological Mott insulators\cite{Pesin10}, and 
Weyl semi-metal (WSM) states\cite{Wan11}, have been proposed\cite{Pesin10, Wan11, Kim12, Go12}.
In the related spinel compounds A$_2$Os$_4$O$_6$, an 
Axion insulator (AI) \cite{Qi08,Essin09,Wan11,Hughes11,Turner12} with bulk magnetic order and  large magnetoelectric 
effect, has been proposed\cite{Wan12}. 

The current theoretical analyses of pyrochlore iridates have focused \emph{either} on the Ir subsystem\cite{Pesin10,Wan11,Kim12,Krempa12,Go12}, 
\emph{or} (excluding non-magnetic R = Y, Eu) on the magnetic moments formed by localized R$^{3+}$ $f$-electrons\cite{Gardner10,Onoda10,Onoda11,Savary12,Lee12}.  
However, as we argue below, coupling between the local moments and the Ir$^{4+}$ $d$-electrons may be important in some compounds, 
but remains largely unexplored (see \cite{Jaubert12,Udagawa12,Ishizuka12} for related prior work).  More generally, most work on correlated, 
strong SOC materials is focused on correlations among $d$-electrons.  An alternate route to introduce correlation into a strong SOC 
system is to couple strong-SOC itinerant carriers to local magnetic moments, and such systems are certainly deserving of greater attention.  

In this Letter, we consider the effect of $f$-$d$ exchange in pyrochlore iridates.  The $f$-$d$ exchange and the Ir electrons together 
generate a RKKY exchange between the localized moments at the R sites, which induces magnetic ordering on the R subsystem. 
The magnetic order on the R subsystem further modifies the electronic structure on the Ir subsystem, leading to WSM and Axion insulator phases.  
Notably, when $f$-$d$ exchange is combined with correlation of the $d$-electrons, we find the WSM is stabilized over a much wider region 
of parameter space than was found for $d$-electron correlations alone\cite{Kim12}.  Based in part on this observation, 
we propose that Nd$_2$Ir$_2$O$_7$ is a candidate to realize the WSM.

In many pyrochlore iridates, a metal-insulator transition and/or magnetic order occurs at a temperature scale of $\sim100$K.  
Such a scale is presumably too large to be driven by RKKY interaction, and instead is probably set 
by $d$-electron magnetic exchange interaction, so the emphasis on Ir subsystem may be justified. However, in some compounds, 
interesting transport and magnetic properties only occur at lower temperatures.
For instance, in Pr$_2$Ir$_2$O$_7$, a chiral spin liquid phase has been proposed  
at $0.3\text{K}<T<1.5\text{K}$ to account for the anomalous hall effect observed 
in this temperature window, where no clear signature of magnetic order is observed\cite{Machida10}.  
In Nd$_2$Ir$_2$O$_7$, the metal-insulator transition, which seems to be associated with magnetic order, occurs at 36K and can be suppressed
by the application of pressure\cite{matsuhira07, Sakata11, Tomiyasu12}.  It should be noted that other studies of the Nd compound show 
different behavior, apparently due to differences in sample preparation\cite{Yanagishima01, Disseler12}. 

We now describe our theoretical model for R$_2$Ir$_2$O$_7$. The R and Ir sites each form a pyrochlore lattice of corner-sharing tetrahedra. 
The non-Kramers R$^{3+}$ ions (R = Pr, Tb, Ho) have an even number of $f$-electrons, 
while the Kramers ions (R = Nd, Sm, Gd, Dy, Yb) have an odd number of $f$-electrons.  
The large SOC of $f$-electrons then leads to a local magnetic moment with integer (non-Kramers) or half-odd-integer (Kramers) total angular momentum $J$.  
The $(2J+1)$-fold degeneracy is then split by the $D_{3d}$ crystal field at the R-site. 
In the Kramers case, this leads to a doublet ground state. In the non-Kramers case, 
the crystal field splits the angular momentum multiplet into doublets and non-magnetic singlets.  A non-Kramers doublet is the 
ground state for the R = Pr iridate\cite{Onoda10,Onoda11}, as well as for isostructural insulating compounds like Tb$_2$Ti$_2$O$_7$ and 
Ho$_2$Ti$_2$O$_7$\cite{Gaulin11,Harris97}, so we assume a doublet ground state.  We ignore effects of higher crystal field levels, which is 
valid for sufficiently large energy gap between the crystal field ground state and first excited state. This is reasonable for the 
R = Pr, Nd cases of greatest interest, where the reported gaps are 168K and 300K, respectively\cite{Machida06,watahiki11}.  
The R-site moment is thus described in all cases by an effective spin-1/2 pseudospin $\boldsymbol{\tau}$.
The $d$-electrons of the Ir subsystem are more itinerant.  Due to the strong SOC, we approximate the Ir subsystem as 
a pyrochlore lattice system with one $j_{\text{eff}} = 1/2$ doublet electron per Ir site\cite{Kim12}. 

\begin{figure}[htp]
\includegraphics[width=7cm]{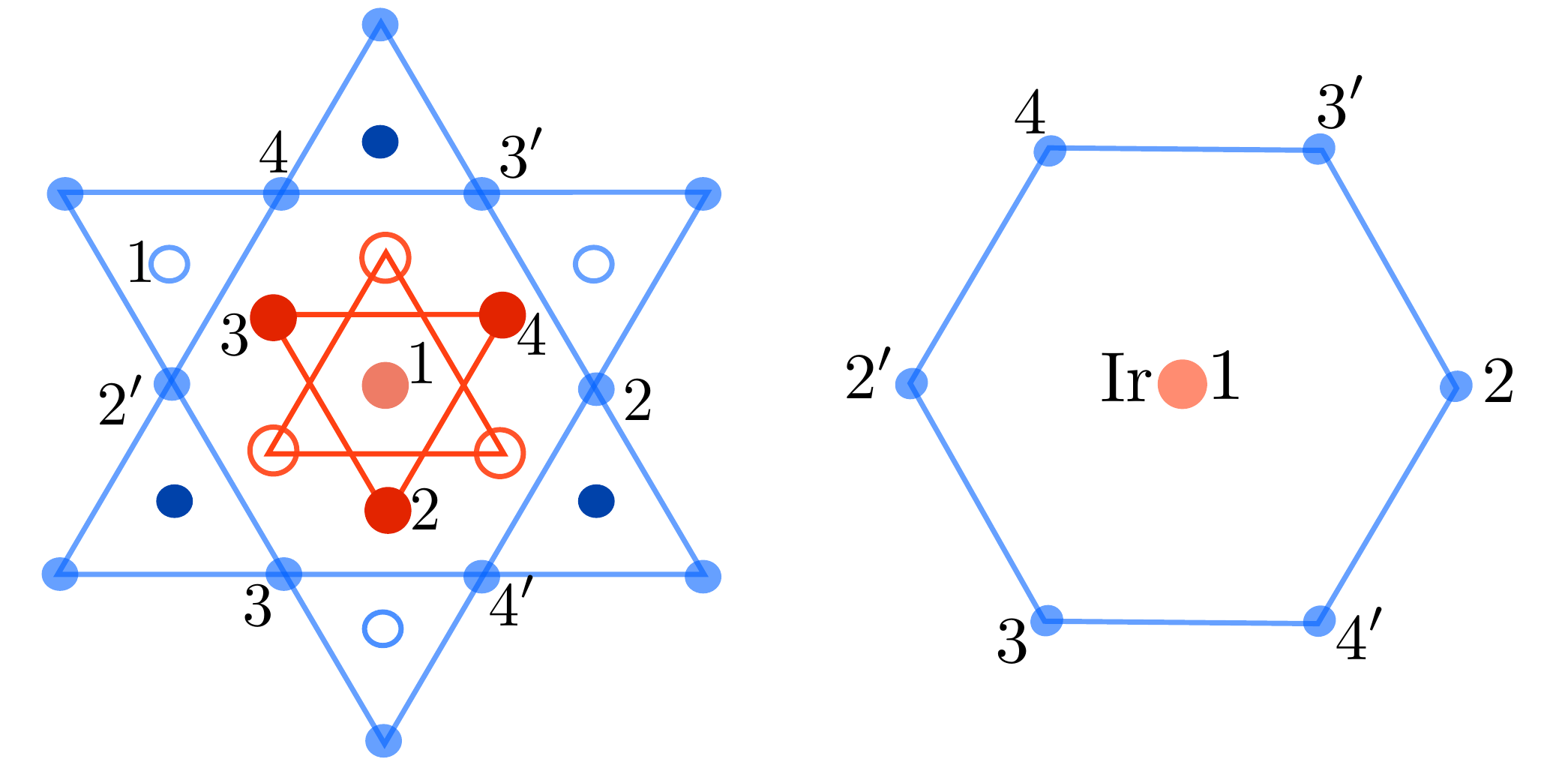}
\caption{(Color online). A projective view of R$_2$Ir$_2$O$_7$ in the (111) plane.  
Left: the neighboring Ir (in dark red) and R (in light blue) tetrahedra.  ``1,2,3,4'' label 
the four sublattices. Ir/R atoms are marked with big/small (red/blue) circles.
Empty/Dark/Light circles indicate that the atoms are below/above/in the (111) plane.
Right: an IrR$_6$ complex singled out from the left figure.} 
\label{fig:fig1}
\end{figure}

For the exchange coupling between R pseudospin $\boldsymbol{\tau}$ and Ir effective spin ${\bf j}$, we invoke 
a general symmetry analysis, beginning with the non-Kramers case.
Under time reversal, $\tau^z \rightarrow -\tau^z, \tau^{x,y} \rightarrow \tau^{x,y}$.  (This is so because $\tau^z$ originates from $J^z$, the component of angular momentum along the local 3-fold axis at the R-site, while $\tau^{x,y}$ originate from $(J^{\pm})^{2 J}$, where $J$ is the total angular momentum.)  On the other hand, under time reversal  ${\bf j} \rightarrow -{\bf j}$.  
This property leads to a remarkable simplification of the coupling -- only $\tau^z$ couples to the Ir spin ${\bf j}$. 
We consider the nearest-neighbor (NN) R-Ir exchange, which, due to space group symmetry, is parametrized by two couplings $c_1, c_2$.  
For the single Ir site labeled Ir1 in Fig.~\ref{fig:fig1}, the $f$-$d$ exchange is
\begin{eqnarray}
{\mathcal H}_{fd} &=&
[c_1 \tau^z_4 -c_2 (\tau^z_2 + \tau^z_3)] j^x_{1}
+ [ c_1 \tau^z_3   - c_2 (\tau_2^z+\tau^z_4)] j^y_{1}
\nonumber \\
&+& 
[ c_1 \tau^z_2   - c_2 (\tau_3^z + \tau^z_4) ]  j^z_{1}
+ [ 2\leftrightarrow 2',3\leftrightarrow 3',4\leftrightarrow 4'],
\nonumber \\
\label{eq:kondo}
\end{eqnarray}
where the labeling of sites is given in Fig.~\ref{fig:fig1}.  
Further details are given in Appendix.~\ref{appendixA}.

In the Kramers case, the Ising part of the $f$-$d$ exchange (coupling of $\tau^z$ to ${\bf j}$) is identical.  
Transverse exchange involving $\tau^{x,y}$ is also permitted by time-reversal symmetry.
Even in the non-Kramers case, while $\tau^{x,y}$ does not couple to the Ir effective spin, it can couple to the Ir charge density.  
In both cases we ignore these transverse couplings, both for simplicity, and because they may be suppressed by 
strong easy-axis anisotropy of the $f$-moments along local $[111]$ axes, which is known to be present in R=Dy, Ho pyrochlore oxides \cite{Gardner10}, and may be present more broadly.
However, effects of transverse exchange may be important, and will be an interesting topic for future study.

For the Ir subsystem, we follow Ref.~\onlinecite{Kim12} and include both the indirect hopping 
of $5d$ electrons through oxygen, and direct hopping between Ir sites, using the following
 Hubbard model,
\begin{equation}
{\mathcal H}_{\text{Ir}} =  \sum_{\langle r r' \rangle}  ( {\mathcal T}^d_{r r' ,\alpha \beta}  
                     +  {\mathcal T}^{id}_{r r',\alpha \beta} ) d^{\dagger}_{r \alpha} d_{r' \beta}  
                     + U \sum_r n_{r,\uparrow} n_{r,\downarrow},
\label{eq:Ir}
\end{equation}
where $d_{r,\alpha}^{\dagger}$ is the electron creation operator,
with $\alpha = \uparrow,\downarrow$ labeling the effective spin $j^z=1/2, -1/2$ states at site $r$,
 and $n_{r,\alpha} = d_{r,\alpha}^{\dagger} d^{\vphantom\dagger}_{r, \alpha}$.  The sum is over NN pairs of Ir sites.
The direct hoppings (${\mathcal T}^d_{r r'}$) involve two parameters\cite{Kim12}, $t_{\sigma}$ and $t_{\pi}$, 
that describe the $\sigma$ and $\pi$ bonding, respectively. To be specific, we follow Ref.~\onlinecite{Kim12} and set
$t_{\pi}= -\frac{2}{3} t_{\sigma}$ throughout the paper. The indirect hopping (${\mathcal T}^{id}_{r r'}$) only has one hopping
parameter which we denote as $t$\cite{Pesin10}. 

The R local moments can couple to each other either via superexchange
through intermediate atoms, by dipole-dipole interaction, or by the RKKY (Ruderman-Kittel-Kasuya-Yosida) exchange
mediated by Ir electrons. Dipole-dipole interactions may play an important role 
for R (= Gd, Tb, Dy, Ho) where a large local magnetic moment is observed\cite{Yanagishima01}. 
RKKY exchange is likely to be the dominant exchange for the other compounds,
as the Curie-Weiss temperatures in many of the isostructural insulating materials R$_2$Sn$_2$O$_7$\cite{Matsuhira02} 
are of much lower magnitude than the correponding iridates. For example,
The Curie-Weiss temperatures $\Theta_{CW}$ are $ -0.35$K in Pr$_2$Sn$_2$O$_7$ and
$ -10$K\cite{Yanagishima01} or $-20$K\cite{Nakatsuji06} in Pr$_2$Ir$_2$O$_7$.  
For the R = Nd compounds, $\Theta_{CW} \approx -0.31$K in the stannate\cite{Matsuhira02} and 
$\Theta_{CW} \approx -19$K in the iridate\cite{Yanagishima01}.

From the above analysis, we obtain our minimal model for R$_2$Ir$_2$O$_7$, which includes the R-Ir 
exchange coupling in Eq.~\eqref{eq:kondo} and the Ir-Ir hopping and interaction Eq.~\eqref{eq:Ir},
${\mathcal H}_{\text{min}} = {\mathcal H}_{fd} + {\mathcal H}_{\text{Ir}}$.

To analyze the phase diagram, we start with the tight-binding model of the Ir subsystem. Following Ref.~\onlinecite{Kim12},
a semi-metal phase is obtained for $-1.67t \lesssim t_{\sigma} \lesssim -0.67t$.  Otherwise, a strong topological band insulator 
(STI) with topological class (1;000) is obtained\cite{Moore07, Fu07, Roy09}.
In the semi-metal phase, at the $\Gamma$ point there is a quadratic band touching (protected by cubic symmetry) at the Fermi energy ($E_F$).  
There are also non-dispersing bands at $E_F$ along the $\Gamma$-L lines; this feature is a consequence of fine-tuning; 
it can be removed by adding a weak next-nearest-neighbor hopping ($t'$) \cite{Kim12}.  
The low-energy features of the band structure agree rather well with the first principles calculation for Y$_2$Ir$_2$O$_7$\cite{Wan11, wanpc}, 
with the differences that the quadratic $\Gamma$-point touching is below $E_F$ and the $\Gamma$-L lines have a small dispersion.

\begin{figure}[htp]
\includegraphics[width=8.5cm]{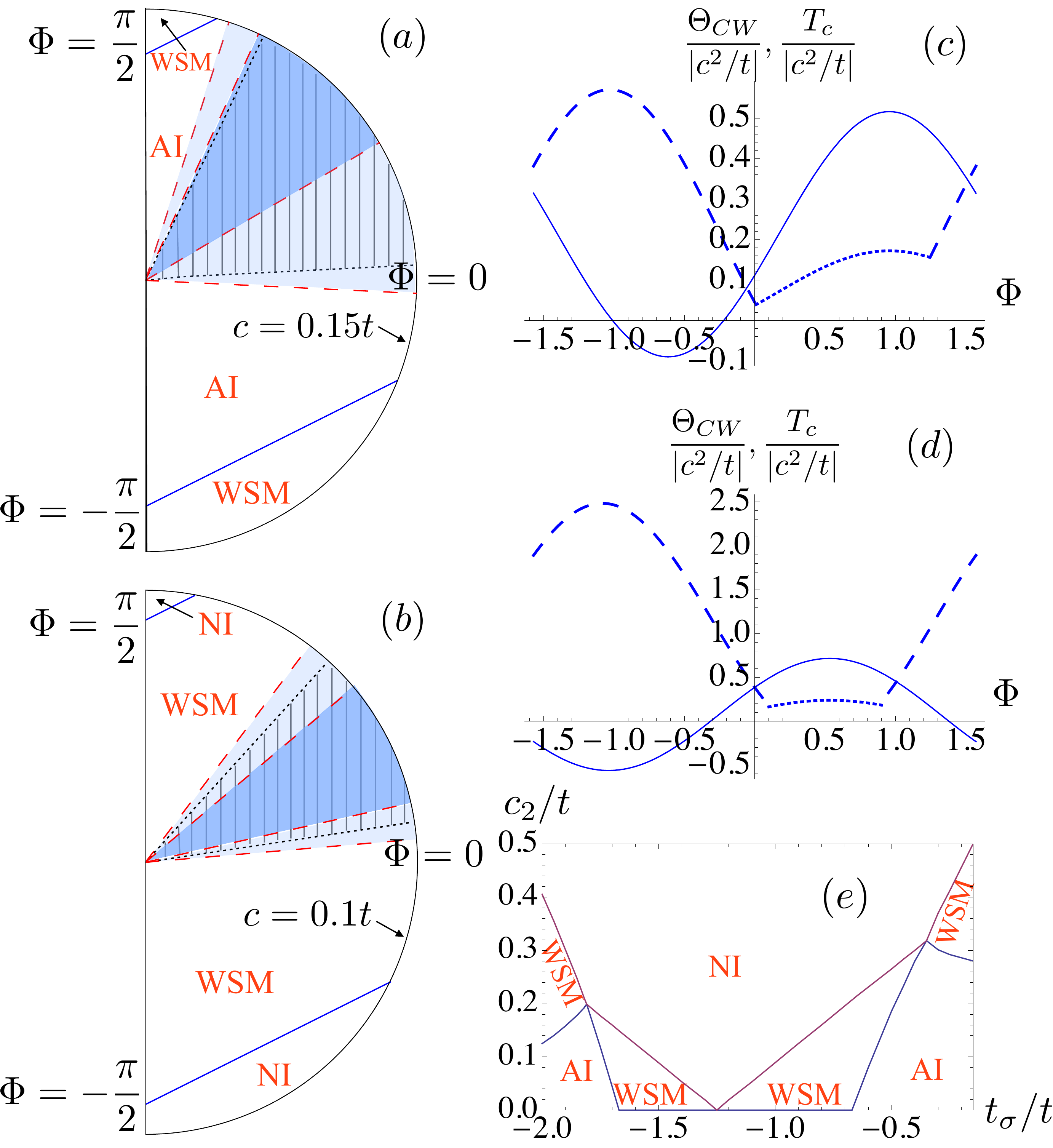}
\caption{(Color online). Phase diagrams of the STI with $t_{\sigma} = -2 t$ (in $a$) and the semi-metal phase with $t_{\sigma}=-t$ (in $b$) 
after including $f$-$d$ exchange. See text for further discussion of the phase diagrams.
In ($c$, $d$), the Curie-Weiss temperature $\Theta_{CW}$ (solid line) and the mean-field ordering temperature $T_c$ (dashed/dotted line) 
of the truncated exchange model (up to 4th neighbor for R system) are plotted against $\Phi$. $T_c$ is obtained by restricting to the 
${\bf q} = {\bf 0}$ magnetic ordering (see Appendix.~\ref{appendixB}). $T_c$ for the all-in all-out (two-in two-out) state
is the dashed (dotted) curve. In ($e$), we plot the $c_2$-$t_{\sigma}$ phase diagram for Ir subsystem where the magnetic order is all-in all-out.   
In the figure, NI = normal insulator.}
\label{fig:fig2}
\end{figure}

Due to the Ising form of the coupling, the model with the $f$-$d$ exchange does not contain 
quantum fluctuations of the $f$-moments, and reduces to a free fermion problem for any fixed configuration of localized moments.  
Finding the ground state amounts to finding the minimum-energy configuration of local moments.  Moreover, certainly $c_1, c_2 \ll t$, 
so the $f$-$d$ exchange can be treated perturbatively, and the leading effect is to generate a RKKY exchange between the $f$-moments.  As shown in Appendix.~\ref{appendixB},
we find that beyond 4th neighbors the RKKY exchange becomes significantly smaller, so we keep only up through 4th-neighbor exchange.
Using the Luttinger-Tizsa method\cite{Luttinger46}, we find that the ground state of the truncated RKKY exchange has a ${\bf q} ={\bf 0}$ 
magnetic order except in the light shaded regions of Fig.~\ref{fig:fig2}($a$,$b$). 
In the light shaded regions, the hard-spin constraint cannot be satisfied, and the nature of the ground state is not presently clear.  
However, it is likely that the ${\bf q} ={\bf 0}$ magnetic order extends at least somewhat into to the light shaded regions. 

Without losing any generality, we can simply focus on the case with $c_1>0$ and define $\Phi \equiv \tan^{-1} (c_2 / c_1)$ and 
$c \equiv \sqrt{c_1^2 + c_2^2}$. As shown in Fig.~\ref{fig:fig2}($a,b$),  for most of parameter space, ``all-in all-out'' magnetic order is favored, 
where every tetrahedron of neighboring R sites has either all $\tau^z$ pointing  in (\emph{i.e.} toward the tetrahedron center), 
or all pointing out.  In the dark shaded region, ${\bf q} = 0$ ``two-in two-out'' magnetic order is obtained, where on every tetrahedron, 
two $\tau^z$ point in and two point out.  [The ${\bf q} = 0$ two-in two-out state also has lowest energy, at least among ${\bf q} = 0$ states, 
in the vertically hatched regions.] Since no ferromagnetic state is observed in any R$_2$Ir$_2$O$_7$, we restrict our discussion to all-in all-out state.
Such order of the R subsystem also induces all-in all-out magnetic order in the Ir subsystem via the $f$-$d$ exchange, 
which acts as a local magnetic field modifying the Ir band structure.
  
As shown in Fig.~\ref{fig:fig2}($a$,$b$), with an $f$-$d$ exchange, the STI phase is converted into an AI for small $c/t$ and then into a WSM.
The semi-metal phase immediately becomes a WSM.  In this case, it is necessary to add a very small $t'$ to stabilize the WSM \cite{Kim12}; 
otherwise, the flat $\Gamma$-L lines of the semi-metal remain at $E_F$.

Returning to the STI case, when time-reversal symmetry is broken but inversion symmetry is preserved, as for the all-in all-out state, 
the magnetoelectric response parameter $\theta$ is still quantized. We can express $\theta$ in terms of the number $n_o({\bf k})$ 
of filled odd parity states at the time reversal invariant momenta (TRIMs) \cite{Turner12,Hughes11}, 
$\frac{\theta}{\pi}  \equiv \frac{1}{2} \sum_{{\bf k} \in \text{TRIMs}} n_{o} ({\bf k}) $ ($\text{mod} \,\, 2$).

\begin{figure}[htp]
\centering
 \includegraphics[width=8cm]{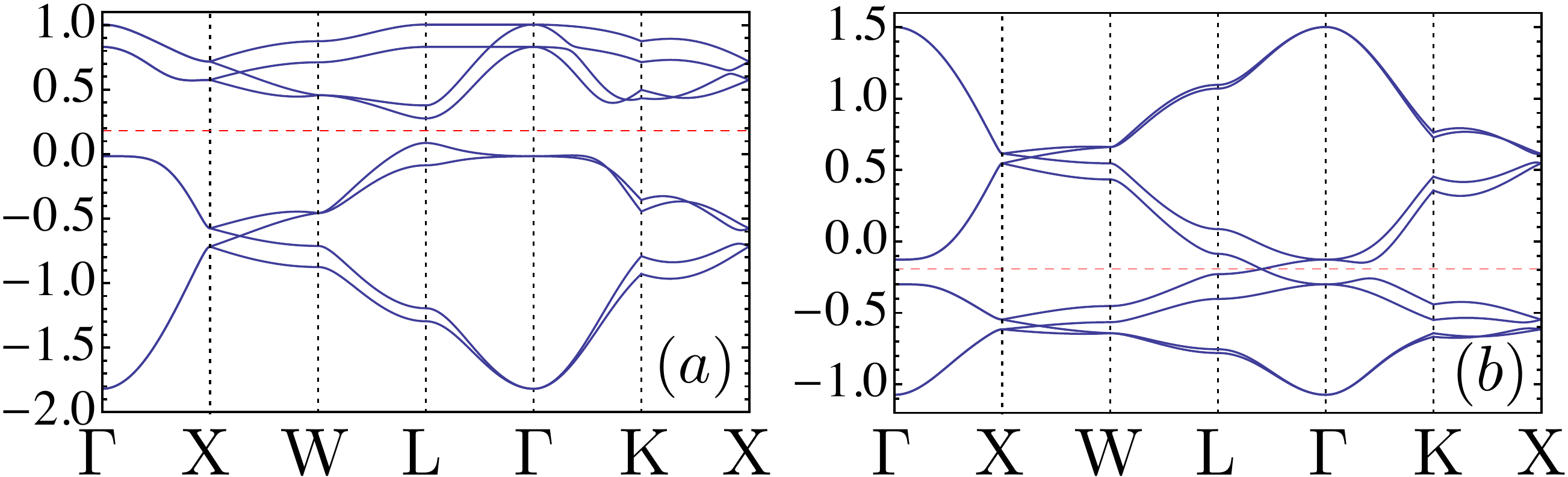} 
\caption{(Color online). The Ir electron band structure with the all-in all-out magnetic order. 
The energy unit is set to $t$ and $c_1=0, c_2=0.05t$. The dashed (red) line is $E_F$.
($a$) is AI phase with $t_{\sigma} = -2t$.  ($b$) is WSM phase with $t_{\sigma} = -t$\cite{nnnhop}.}
\label{fig:fig3}
\end{figure}

As shown in Fig.~\ref{fig:fig2}($a$), a region of AI phase with $\theta = \pi$ is obtained in the vicinity of the STI phase
and the band structure of an AI induced by $f$-$d$ exchange is shown in Fig.~\ref{fig:fig3}($a$). 

The Fermi surface of WSM is composed of the Weyl points, around which the spectrum is linear and gapless. 
We locate the Weyl points explicitly by examining the spectrum and density of states, which shows the 
characteristic $(E-E_F)^2$ scaling. As $c$ increases, the 8 Weyl points of the WSM in Fig.~\ref{fig:fig2}($b$) are created at 
$\Gamma$ and annihilated at the L points, and lie along the $\Gamma$-L lines [Fig.~\ref{fig:fig3}(b)].

\begin{figure}[htp]
\includegraphics[width=6cm]{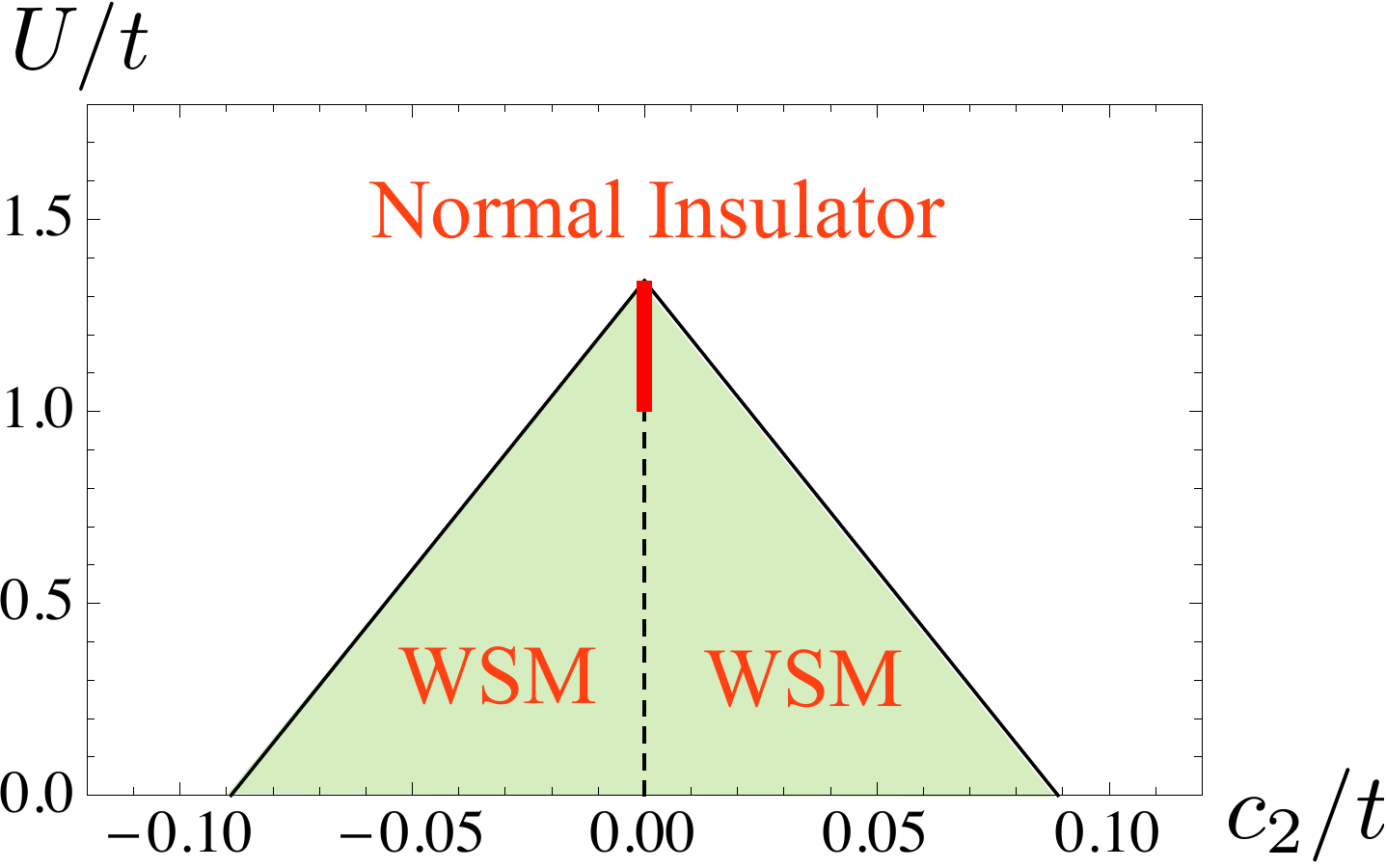}
\caption{(Color online). The mean field phase diagram at $t_{\sigma} = -t$ and $c_1 =0$. The (red) thick line is the narrow WSM phase induced
by Ir correlation. The dashed line is the nonmagnetic Ir semi-metal phase. The (green) region is the broad WSM phase after including $f$-$d$ exchange. 
All-in all-out magnetic order is present throughout the phase diagram. }
\label{fig:fig6}
\end{figure}

We now include the correlation on the Ir subsystem. Without the $f$-$d$ exchange, the Hartree mean-field
analysis of the extended Hubbard model on Ir subsystem gives a rather narrow region of WSM phase\cite{Kim12}. 
With $f$-$d$ exchange, the RKKY exchange induces magnetic order even with weak correlations and thus gives a
WSM phase. We start from the metallic phase and include both the Hubbard interaction (via the same mean-field approach used in Ref.~\onlinecite{Kim12}) and $f$-$d$ exchange. 
Since both correlation and RKKY interaction alone give an all-in all-out magnetic order on Ir system, we restrict attention to the all-in all-out state.  We decouple the Hubbard interaction, 
\begin{equation}
	U n_{i\uparrow} n_{i\downarrow} \to -\frac{2U}{3} {\bf j}_i^2
                                                    \to -\frac{4U}{3} \langle {\bf j}_i \rangle\cdot {\bf j}_i + \frac{2U}{3}\langle {\bf j}_i \rangle^2,
\end{equation}
and find a broad region of WSM phase (see Fig.~\ref{fig:fig6}).

A much-studied compound in the R$_2$Ir$_2$O$_7$ series is R = Pr, which exhibits a metallic ground state 
and anomalous Hall effect without any observable magnetic order for $0.3\text{K} <T < 1.5\text{K}$\cite{Nakatsuji06,Machida07,Machida10}. 
The Pr moments freeze at $T< 0.3\text{K}$. Our theoretical analysis focuses on the all-in all-out region of
the phase diagrams in Fig.~\ref{fig:fig2}. Pr$_2$Ir$_2$O$_7$ may be located in or near the light shaded region of Fig.~\ref{fig:fig2}($b$), 
where the RKKY exchange may be more frustrated.
Moreover, quantum fluctuations may be important for Pr$_2$Ir$_2$O$_7$, perhaps originating from the transverse part of the $f$-$d$ exchange, 
from superexchange between the $f$-moments\cite{Onoda10}, or both.  Such quantum fluctuations may suppress the magnetic order, 
perhaps cooperating with classical frustration.

Nd$_2$Ir$_2$O$_7$ has been observed to order magnetically below the metal-insulator transition at $T_c=36$K\cite{matsuhira07, watahiki11,Sakata11,Tomiyasu12}.  
Recent neutron scattering experiments\cite{Tomiyasu12} on Nd$_2$Ir$_2$O$_7$ suggest an all-in all-out spin configuration for both Nd and Ir systems.  
The data was interpreted in terms of a static local field setting in for the Nd moments immediately upon cooling through $T_c$, 
while Nd magnetic order is not detected until 15K.  Another experiment\cite{Sakata11} finds that applying pressure to Nd$_2$Ir$_2$O$_7$ suppresses the metal-insulator
transition and leads to a metallic state that exhibits a negative magnetoresistance.  
Since $\Theta_{CW} = -19$K in Nd$_2$Ir$_2$O$_7$, according to Fig.~\ref{fig:fig2}($c$,$d$) 
we expect Nd$_2$Ir$_2$O$_7$ to be in the region with $\Phi<0$ in order to have an antiferromagnetic $\Theta_{CW}$. 
Moreover, $T_c > |\Theta_{CW}|$ and the all-in all-out ordering are also consistent with Fig.~\ref{fig:fig2}($c$,$d$). 
The metal-insulator transition and magnetic order are likely driven by a combination of RKKY exchange and Ir magnetic exchange.  
Indeed, Ref.~\onlinecite{Tomiyasu12} interpreted their results in terms of Ir magnetic order setting in at $T_c$, with Nd order setting in only at 
15K (driven by $f$-$d$ exchange) due to the presumably small Ir ordered moment.  It would be interesting to see whether our model 
can indeed produce such behavior at finite temperature.  Assuming all-in all-out ordering is indeed present, the Ir electrons may 
realize a WSM phase within the magnetically ordered state.  Indeed, the transport properties of the WSM are close to those of an 
insulator\cite{Hosur12}, and appear consistent with the resistivity of the Nd compound below $T_c$.  
Applying pressure to Nd$_2$Ir$_2$O$_7$\cite{Sakata11} may modify the Ir electron hopping, generating a metallic state.

To summarize, we have constructed and analyzed a minimal model accounting for $f$-$d$ exchange
in rare-earth-based pyrochlore iridates R$_2$Ir$_2$O$_7$. With this model we find broad regions of 
Axion insulator and Weyl semi-metal phases.


\emph{Acknowledgements} -- We are grateful for helpful discussion with L. Balents, A. Essin, Y.-B. Kim, S. Onoda, 
Y. Ran, Fa Wang, and W. Witczak-Krempa, and useful correspondence with X. Wan.  This work is supported by DOE 
award no. DE-SC0003910.  
Some of this work was carried out at the Aspen Center for Physics and the Kavli Institute for Theoretical Physics; 
our stays there were supported in part by NSF grant no. 1066293, and NSF grant no. PHY11-25915, respectively.

\appendix

\section{$f$-$d$ exchange between the R moment and Ir moment}
\label{appendixA}

Both Ir and R sublattices are pyrochlore lattices, and can be viewed as FCC lattices with a four-site basis.  
We use the BCC primitive vectors
\begin{eqnarray}
{\bf b}_1 & = & (0,\frac{1}{2},\frac{1}{2}), \\
{\bf b}_2 & = & (\frac{1}{2},0,\frac{1}{2}), \\
{\bf b}_3 & = & (\frac{1}{2},\frac{1}{2},0). 
\end{eqnarray}
For the Ir pyrochlore lattice,
we choose the following reference points for four sublattices,
\begin{eqnarray}
\text{Ir}_1 & = & \frac{1}{4}(0,0,0), \\
\text{Ir}_2 & = & \frac{1}{4}(0,1,1), \\
\text{Ir}_3 & = & \frac{1}{4}(1,0,1), \\
\text{Ir}_4 & = & \frac{1}{4}(1,1,0).
\end{eqnarray}
For the R pyrochlore lattice, we choose 
\begin{eqnarray}
\text{R}_1 & = & (0,\frac{1}{2},0), \\ 
\text{R}_2 & = & (0,\frac{3}{4},\frac{1}{4}), \\
\text{R}_3 & = & (\frac{1}{4},\frac{1}{2},\frac{1}{4}), \\
\text{R}_4 & = & (\frac{1}{4},\frac{3}{4},0). 
\end{eqnarray}

\begin{widetext}
The symmetry-allowed
 $f$-$d$ exchange between the R moment and the Ir moment can be written as 
\begin{eqnarray}
\label{eq:rkky}
{\mathcal H}_{fd} & = & \sum_{\bf r} \tau^z_1 ({\bf r}) 
\Big[ c_1 \big[j^x_{2}({\bf r}) + j^x_{2} ({\bf r}-{\bf b}_2+{\bf b}_3)  
+ j^y_{3} ({\bf r} - {\bf b}_2 + {\bf b}_3)+ j^y_{3} ({\bf r} + {\bf b}_1 - {\bf b}_2)
+ j^z_{4} ({\bf r}) 
\nonumber \\
&+&   j^z_{4} ({\bf r} +{\bf b}_1 - {\bf b}_2) \big]  +
c_2  \big[j^y_{2}({\bf r}) + j^y_{2} ({\bf r}-{\bf b}_2+{\bf b}_3) +j^z_{2}({\bf r}) + j^z_{2} ({\bf r}-{\bf b}_2+{\bf b}_3)  
+ j^x_{3} ({\bf r} - {\bf b}_2 + {\bf b}_3)
\nonumber \\
&+&  j^x_{3} ({\bf r} + {\bf b}_1 - {\bf b}_2)+  j^z_{3} ({\bf r} - {\bf b}_2 + {\bf b}_3)+ j^z_{3} ({\bf r} + {\bf b}_1 - {\bf b}_2)
+ j^x_{4} ({\bf r}) + j^x_{4} ({\bf r} +{\bf b}_1 - {\bf b}_2) + j^y_{4} ({\bf r}) 
\nonumber \\
&+&  j^y_{4} ({\bf r} +{\bf b}_1 - {\bf b}_2)  \big]  \Big]
\nonumber \\
&+& \tau^z_2 ({\bf r}) 
\Big[ c_1 \big[j^x_{1}({\bf r}+{\bf b}_1) 
+ j^x_{1} ({\bf r}+{\bf b}_1-{\bf b}_2+{\bf b}_3)  - j^y_{4} ({\bf r} + {\bf b}_1)
- j^y_{4} ({\bf r} + {\bf b}_1 - {\bf b}_2 ) 
\nonumber \\
&-& j^z_{3} ({\bf r} + {\bf b}_1 - {\bf b}_2 ) 
- j^z_{3} ({\bf r} + {\bf b}_1 - {\bf b}_2+{\bf b}_3) \big]  +c_2 \big[ -j^y_{1}({\bf r}+{\bf b}_1)- j^y_{1} ({\bf r}+{\bf b}_1-{\bf b}_2+{\bf b}_3)  - j^z_{1}({\bf r}+{\bf b}_1) 
\nonumber \\
&-& j^z_{1} ({\bf r}+{\bf b}_1-{\bf b}_2+{\bf b}_3) 
+ j^x_{4} ({\bf r} + {\bf b}_1 - {\bf b}_2 ) + j^x_{4} ({\bf r} + {\bf b}_1)- j^z_{4} ({\bf r} + {\bf b}_1 - {\bf b}_2 )- j^z_{4} ({\bf r} + {\bf b}_1)
\nonumber \\
&+&j^x_{3} ({\bf r} + {\bf b}_1 - {\bf b}_2 ) +
 j^x_{3} ({\bf r} +{\bf b}_1 - {\bf b}_2+{\bf b}_3 ) - j^y_{3} ({\bf r} + {\bf b}_1 - {\bf b}_2 ) 
- j^y_{3} ({\bf r} +{\bf b}_1 - {\bf b}_2+{\bf b}_3) \big]   \Big] 
\nonumber \\
&+& \tau^z_3 ({\bf r}) \Big[ c_1 \big[
j^y_{1} ({\bf r}+{\bf b}_3) + j^y_{1} ({\bf r}+{\bf b}_1) - j^z_{2}({\bf r}) -j^z_{2}({\bf r}+{\bf b}_3)
-j^x_{4}({\bf r}) - j^x_{4} ({\bf r}+{\bf b}_1) \big]
\nonumber \\
&+& c_2 \big[-j^x_{1} ({\bf r}+{\bf b}_3)  - j^x_{1} ({\bf r}+{\bf b}_1)- j^z_{1} ({\bf r}+{\bf b}_3) - j^z_{1} ({\bf r}+{\bf b}_1) - j^x_{2} ({\bf r}) - j^x_{2} ({\bf r} + {\bf b}_3) 
\nonumber \\
&+& j^y_{2} ({\bf r}) + j^y_{2} ({\bf b}_3) 
+ j_{4}^y({\bf r}) + j^y_{4}({\bf r} + {\bf b}_1)
- j_{4}^z({\bf r})  - j_{4}^z ({\bf r} + {\bf b}_1)
\big]
\Big]
\nonumber \\
&+&  \tau^z_4({\bf r}) \Big[
c_1 \big[  j^z_{1} ({\bf r}+{\bf b}_3) + j^z_{1} ({\bf r} + {\bf b}_1 -{\bf b}_2 + {\bf b}_3) 
- j^y_{2}({\bf r} -{\bf b}_2 + {\bf b}_3) - j^y_{2} ({\bf r}-{\bf b}_2 + {\bf b}_3) 
\nonumber \\
&-& j^x_{3} ({\bf r} - {\bf b}_2 + {\bf b}_3) - j^x_{3} ({\bf r} + {\bf b}_1 - {\bf b}_2 + {\bf b}_3)
 \big]
+c_2 \big[
-j^x_{1} ({\bf r}+{\bf b}_3) - j^x_{1} ({\bf r} + {\bf b}_1 -{\bf b}_2 + {\bf b}_3) 
-j^y_{1} ({\bf r}+{\bf b}_3) 
\nonumber\\
&-& j^y_{1} ({\bf r} + {\bf b}_1 -{\bf b}_2 + {\bf b}_3)- j^x_{2} ({\bf r} -{\bf b}_2 + {\bf b}_3) -j^x_{2} ({\bf r} + {\bf b}_3) 
+j^z_{2} ({\bf r} -{\bf b}_2 + {\bf b}_3) +j^z_{2}({\bf r} + {\bf b}_3) 
\nonumber \\
&-& j^y_{3} ({\bf r} - {\bf b}_2 + {\bf b}_3) 
- j^y_{3} ({\bf r} +{\bf b}_1 -{\bf b}_2 + {\bf b}_3)
+j^z_{3} ({\bf r} - {\bf b}_2 + {\bf b}_3) + j^z_{3} ({\bf r} + {\bf b}_1 - {\bf b}_2 + {\bf b}_3)
 \big]
\Big] \text{.}
\end{eqnarray}
\end{widetext}
Here $\tau^z_i( {\bf r} )$ is the pseudospin for the R site in unit cell labeled by ${\bf r}$ with sublattice index $i = 1,\dots,4$.  (Recall that $\tau^z$ is the component of pseudospin along the local 3-fold axis at the R site.)  Similarly, $j^{\mu}_i({\bf r})$ is the $\mu = x,y,z$ component of the Ir effective spin (in global cubic coordinates), for the Ir site in unit cell ${\bf r}$ at sublattice $i$.
$c_1$ and $c_2$ are the two parameters allowed by space group symmetries. 

It is  convenient below to work in a local coordinate system for the Ir effective moment. So,
we transform the Ir effective spin from glocal to local IrO$_6$ octahedral coordinate system by
\begin{equation}
j^{\mu}_{i} ({\bf r}) = {\mathcal R}_i^{\mu\nu} j^{\nu}_{i,L} ({\bf r}),
\end{equation}
where ${\bf j}_{i,L} ({\bf r})$ is the Ir effective spin in the local IrO$_6$ octahedral coordinate
system.  The transformation matrices for four sublattices are
\begin{eqnarray}
{\mathcal R}_1 & = & \left[\begin{array}{ccc}
2/3& -1/3& -2/3 \\
-1/3& 2/3& -2/3 \\
2/3 & 2/3 &1/3
\end{array}
\right] \text{,} \\
{\mathcal R}_2 &=& \left[\begin{array}{ccc}
2/3& 2/3& 1/3 \\
-2/3& 1/3& 2/3 \\
1/3 & -2/3 &2/3
\end{array}
\right]  \text{,} \\
{\mathcal R}_3 &=& \left[\begin{array}{ccc}
1/3& -2/3& 2/3 \\
2/3& 2/3& 1/3 \\
-2/3 & 1/3 & 2/3
\end{array}
\right] \text{,} \\
{\mathcal R}_4 &=& \left[\begin{array}{ccc}
1/3& -2/3& 2/3 \\
-2/3& -2/3& -1/3 \\
2/3 & -1/3 & -2/3
\end{array}
\right]  \text{.}
\end{eqnarray}

\section{RKKY interaction between the R moments}
\label{appendixB}

We consider the  Hamiltonian
\begin{equation}
{\mathcal H} = {\mathcal H}_{\text{Ir}}  + {\mathcal H}_{fd},
\end{equation}
where $ {\mathcal H}_{\text{Ir}} $ is the tight-binding Hamiltonian without the Hubbard interaction,
\begin{equation}
{\mathcal H}_{\text{Ir}} = \sum_{\langle r r' \rangle} ({\mathcal T}_{r r',\alpha\beta}^{d}
+{\mathcal T}_{r r',\alpha\beta}^{id}  ) d^{\dagger}_{r \alpha}d_{r'\beta}.
\end{equation} 
The $f$-$d$ exchange ${\mathcal H}_{fd}$ can be written compactly as
\begin{equation}
{\mathcal H}_{fd} = \sum_{ {\bf r},{\bf r}'  } \sum_{ij} \tau^z_i ({\bf r}) 
f_{ij}^{\mu} ({\bf r} - {\bf r}') j^{\mu}_{j} ({\bf r}') \text{.}
\end{equation}
Here, and below, sums on repeated indices are implied.  The exchange is characterized by the function $f^{\mu}_{ij}({\bf r})$, which can be read off from Eq.~\eqref{eq:rkky}.

Diagonalizing the tight-binding Hamiltonian ${\mathcal H}_{\text{Ir}}$ gives
\begin{equation}
{\mathcal H}_{\text{Ir}} = \sum_{{\bf k},\lambda} \epsilon_{{\bf k}\lambda} 
c^{\dagger}_{{\bf k}\lambda} c_{{\bf k}\lambda} \text{,}
\end{equation}
where $\lambda$ labels the eight eigenstates for each ${\bf k}$, and 
\begin{equation}
d^{\dagger}_{{\bf k}i\alpha} = \sum_{\lambda} c^{\dagger}_{{\bf k}\lambda} {\mathcal U}^{\vphantom\dagger}_{\lambda,i\alpha} ({\bf k}). 
\end{equation}

We now treat ${\mathcal H}_{fd}$ as a perturbation to the Ir tight-binding Hamiltonian ${\mathcal H}_{\text{Ir}}$. 
By standard second-order perturbation theory, we obtain
the effective RKKY exchange between the local moments of the R subsystem, 
\begin{equation}
{\mathcal H}_{\text{RKKY}} =
 \sum_i \frac{ \langle \Phi_0 |{\mathcal H}_{fd}| \Phi_i \rangle\langle \Phi_i |{\mathcal H}_{fd}
| \Phi_0 \rangle   }{ E_0 - E_i },
\end{equation}
in which, $|\Phi_0\rangle$ and $E_0$ is the ground state and the corresponding energy,
and $|\Phi_i \rangle $ and $E_i$ is the $i$th excited state and the corresponding energy. 
This result can be written
\begin{equation}
{\mathcal H}_{\text{RKKY}} = \frac{1}{2}\sum_{{\bf r}_1 i_1, {\bf r}_2 i_2} 
J_{i_1 i_2} ({\bf r}_1 - {\bf r}_2) \tau^z_{ i_1}  ({\bf r}_1) \tau^z_{ i_2} ({\bf r}_2) 
\end{equation}
with $J_{i_1 i_2} ({\bf r}_1 - {\bf r}_2) \equiv  \tilde{J}_{i_1 i_2} ({\bf r}_1 - {\bf r}_2) 
+ \tilde{J}_{i_2 i_1} ({\bf r}_2 - {\bf r}_1)$, 
where
\begin{widetext}
\begin{eqnarray}
\tilde{J}_{i_1 i_2} ({\bf r}_1 - {\bf r}_2)& =& \frac{1}{4} \sum_{{\bf r}_1',{\bf r}_2'}
f^{\mu_1}_{i_1j_1} ({\bf r}_1-{\bf r}_1') f^{\mu_2}_{i_2j_2} ({\bf r}_2-{\bf r}_2') 
{\mathcal R}_{j_1}^{\mu_1 \nu_1} {\mathcal R}_{j_2}^{\mu_2 \nu_2} 
\sigma_{\alpha_1\beta_1}^{\nu_1} \sigma_{\alpha_2 \beta_2}^{\nu_2}
\frac{1}{N_c^2}\sum_{{\bf k}_1,{\bf k}_2} e^{i ({\bf k}_2 -{\bf k}_1)\cdot ({\bf r}_1'-{\bf r}_2')}
\nonumber \\
&&
\mathcal{U}_{\lambda_1,j_1\alpha_1} ({\bf k}_1) \mathcal{U}^{\ast}_{\lambda_1,j_2\beta_2}({\bf k}_1)
\mathcal{U}_{\lambda_2,j_2\alpha_2}({\bf k}_2) \mathcal{U}^{\ast}_{\lambda_2,j_1\beta_1}({\bf k}_2) \frac{ \Theta(\epsilon_{F} - \epsilon_{\lambda_1} ({\bf k}_1)) \Theta( \epsilon_{\lambda_2} ({\bf k}_2) - \epsilon_{F})}{  \epsilon_{\lambda_1} ({\bf k}_1) - \epsilon_{\lambda_2} ({\bf k}_2) } \text{.}
\end{eqnarray}
Here,  $\sigma^{\nu}$ is the Pauli matrix, $N_c$ is the number of unit cell, and $\Theta(x)$ the Heaviside step function.

\begin{figure}[htp]
\centering
\subfigure{ \includegraphics[width=12cm]{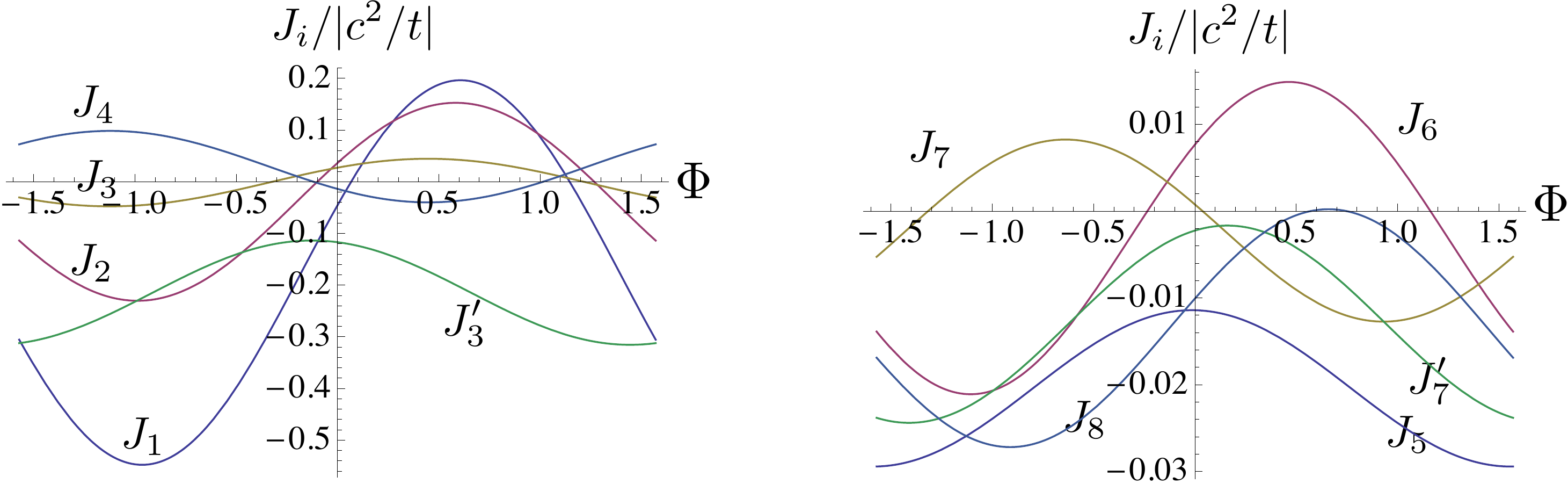} \label{fig:fig1a}}
\subfigure{ \includegraphics[width=12cm]{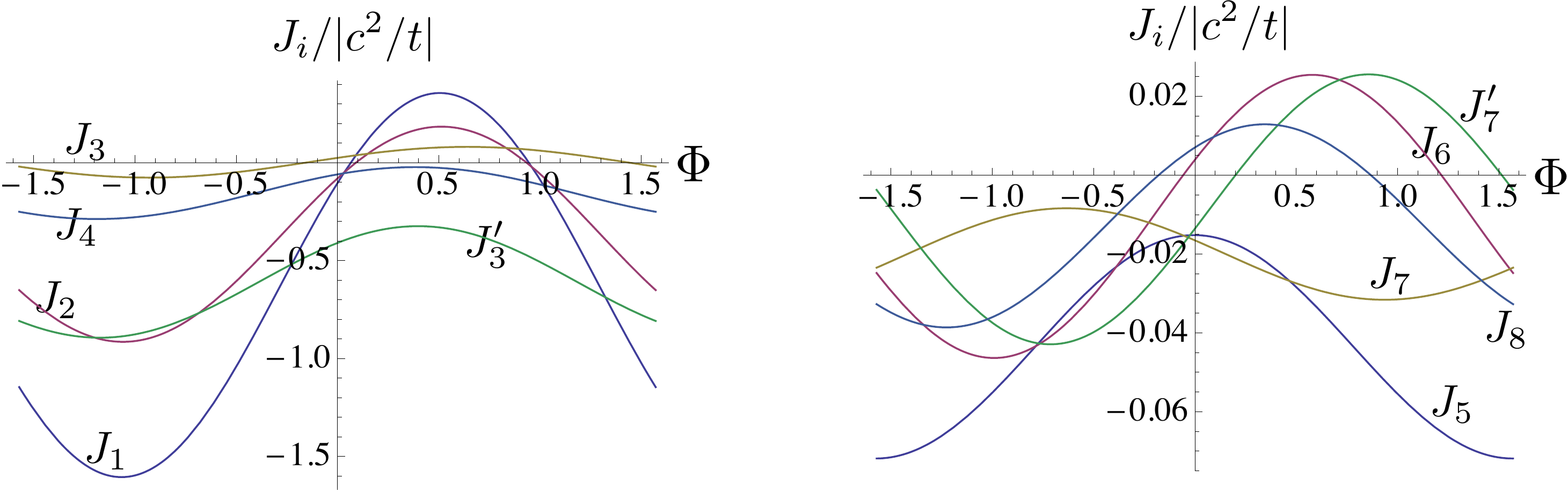} \label{fig:fig1b}}
\caption{(Color online) The dependence of RKKY exchanges on the angle $\Phi$. The results are
obtained for a finite system with $10\times 10 \times 10$ unit cells. 
$J_i$ is the $i$th neighbor exchange.
$J_3$ is the exchange on 3rd neighbor bonds that have one lattice site at the midpoint of the bonds. 
$J_3'$ is the exchange on 3rd neighbor bonds that have no lattice site at the midpoint of the bonds. 
$J_7$ is the exchange on 7th neighbor bonds that have two additional lattice sites along the bonds. $J_7'$ is the exchange on 7th neighbor bonds that have no lattice sites along the bonds. 
Upper: STI phase with $t_{\sigma} =- 2t$.
Lower: semi-metal phase with $t_{\sigma} = -t$.
}
\label{supplefig1}
\end{figure}

Since the RKKY exchanges are quadratic in $c_1$ and $c_2$, we can simply focus on the regime
with $c_1 \geq 0$ and define an angle $\Phi \equiv \tan^{-1} (c_2/c_1)$ and $c \equiv \sqrt{c_1^2 +c_2^2}$.  
As shown in Fig.~\ref{supplefig1}, we compute the RKKY exchanges of a finite system with $10 \times 10 \times 10$ 
unit cells for both the STI phase with $t_{\sigma} =-2t$
and the semi-metal phase with $t_{\sigma}  = -t$. Clearly, the RKKY exchanges beyond
4th neighbor become significantly smaller. We now focus on the truncated RKKY exchanges upto 4th neighbor.  
The Curie-Weiss temperature is found to be,
\begin{equation}
\Theta_{CW} = \frac{1}{2} (J_1 + 2 J_2 - 3 J_3 - 3J_3' + 2 J_4). 
\end{equation}
The transition from the high temperature paramagnetic phase to low temperature ordered phase is continuous at
mean field level. The transition from paramagnetic phase to all-in all-out state is of three dimensional Ising type.  
We can then determine the critical temperatures for this transition 
by the usual condition of marginal stability (vanishing of the quadratic term in the Landau theory) of the free energy. 
Restricting to ${\bf q} = {\bf 0}$ magnetic order, we find that the critical temperatures for all-in all-out and 
two-in two-out states are given by
\begin{eqnarray}
T_c (\text{all-in all-out})     & = & \frac{1}{2} (-J_1 - 2 J_2 - J_3 - J_3' - 2 J_4),
\\
T_c (\text{two-in two-out}) & = & \frac{1}{6} (J_1 + 2 J_2 - 3 J_3 - 3 J_3' + 2 J_4),
\end{eqnarray}
respectively. The actual transition temperature is determined by the higher one of the above transition 
temperatures.

\end{widetext}

\bibliography{fdexchange}

\begin{thebibliography}{10}
\bibitem{nnnhop}%
  \BibitemOpen
  \bibinfo {note} {\textnormal{We add a small second neighbor hopping
  ($t'_{\sigma} = 0.04 t_{\sigma}$ and $t'_{\pi} = 0.04 t_{\pi}$) to break the
  flatness along the $\Gamma$-L lines. }}%
  \bibAnnoteFile{NoStop}{nnnhop}%
\end{thebibliography}

\end{document}